\documentclass[preprint,12pt]{aastex}

\newcommand{\bq}{\begin{equation}}
\newcommand{\eq}{\end{equation}}

\newcommand{\3}{$_3$}
\newcommand{\kms}{km~s$^{-1}$}
\newcommand{\dvi}{\Delta \rm{v}_{int}}
\newcommand{\dvilow}{\Delta \rm{v}_{int,low}}
\newcommand{\dvihigh}{\Delta \rm{v}_{int,high}}
\newcommand{\dvo}{\Delta \rm{v}_{obs}}
\newcommand{\dvoa}{\Delta \rm{v}_{obs}(1,1)}
\newcommand{\dvoc}{\Delta \rm{v}_{obs}(3,3)}
\newcommand{\tm}{\tau_{\rm{m}}}
\newcommand{\ta}{\tm(1,1)}
\newcommand{\talow}{\ta_{\rm{low}}}
\newcommand{\tahigh}{\ta_{\rm{high}}}
\newcommand{\tc}{\tm(3,3)}
\newcommand{\tclow}{\tc_{\rm{low}}}
\newcommand{\tchigh}{\tc_{\rm{high}}}
\newcommand{\inten}{\rm{I}(\rm{v},\sigma,\tm)}
\newcommand{\simgt}{\lower.5ex\hbox{$\; \buildrel > \over \sim \;$}}
\newcommand{\simlt}{\lower.5ex\hbox{$\; \buildrel < \over \sim \;$}}

\begin{document}
 
\title{NH\3 in the Central 10 pc of the Galaxy. II. Determination of Opacity for Gas with Large Linewidths}
 
\author{
Robeson S. McGary\altaffilmark{1} and
Paul T.P. Ho\altaffilmark{1}}

\altaffiltext{1}{Harvard-Smithsonian Center for Astrophysics, 60 Garden Street, Cambridge, MA 02138, \\rmcgary@cfa.harvard.edu, pho@cfa.harvard.edu}


\begin{abstract}
The 23~GHz emission lines from the NH\3 rotation inversion transitions
are widely used to investigate the kinematics and physical conditions
in dense molecular clouds.  The line profile is composed of hyperfine
components which can be used to calculate the opacity of the gas
\citep{ho83}.  For intrinsic linewidths of a few \kms, the 18 magnetic
hyperfine components blend together to form a line profile composed of
five quadrupole hyperfine lines.  If the intrinsic linewidth exceeds
one half of the separation of these quadrupole hyperfine components
($\sim5-10$~\kms) these five lines blend together and the observed
linewidths greatly overestimate the intrinsic linewidths.  If
uncorrected, these artificially broad linewidths will lead to
artificially high opacities.  We have observed this effect in our NH\3
data from the central 10 pc of the Galaxy where uncorrected NH\3(1,1)
linewidths of $\sim30$ \kms ~exaggerate the intrinsic linewidths by
more than a factor of two \citep{gen87}.  Models of the effect of
blending on the line profile enable us to solve for the intrinsic
linewidth and opacity of NH\3 using the observed linewidth and
intensity of two NH\3 rotation inversion transitions.  By using the
observed linewidth instead of the entire line profile, our method may
also be used to correct linewidths in historical data where detailed
information on the shape of the line profile is no longer available.
We present the result of the application of this method to our
Galactic Center data.  We successfully recover the intrinsic linewidth
($\langle\dvi\rangle\approx15$~\kms) and opacity of the gas. Clouds
close to the nucleus in projected distance as well as those that are
being impacted by Sgr A East show the highest intrinsic linewidths.
The cores of the ``southern streamer'' \citep{ho91,coi99,coi00} and
the ``50 \kms'' giant molecular cloud (GMC, \citet{gus81}) have the
highest opacities.

\end{abstract}


\section{Introduction \label{intro}}
The metastable (J=K) rotation inversion transitions of NH\3 are
effective tracers of dense molecular material ($\rm{N_H}\approx10^4 -
10^5$~cm$^{-3}$) \citep{ho83}.  The NH\3 rotation inversion
transitions are most often used to trace dense cores of giant
molecular clouds (GMCs), but they are also observed in the
``streamers'' near the Galactic Center \citep{ho91,coi99,coi00,mcg01}
which will be our example data for this paper.  With frequencies near
23 GHz (1.3 cm), these transitions are easily observable with radio
telescopes such as the Very Large Array\footnote{The National Radio
Astronomy Observatory is a facility of the National Science Foundation
operated under cooperative agreement by Associated Universities, Inc.}
(VLA) with only minimal interference by the atmosphere.  Compared to
high frequency tracers such as the 3~mm transitions of HCN(J=1-0) and
HCO$^+$(J=1-0) which often have $\tau>1$, these NH\3 transitions tend
to be more optically thin, even in the case of low temperatures.  As a
result, spectra of NH\3 rotation inversion transitions are usually
free of self-absorption effects.

The NH\3 rotation inversion line is composed of 18 magnetic hyperfine
lines which are grouped into five main features (a detailed
description of the NH\3 rotation inversion transitions can be found in
\citet{tow75} and \citet{ho83}). The spacing of the magnetic hyperfine
lines within each of the five components is too close to resolve
except in cool and quiescent dense cores where the intrinsic
linewidths of the gas can be less than 1~\kms.  The effect of blending
of the 18 magnetic hyperfine components on the observed linewidth is
discussed in detail in \citet{ho77} and \citet{bar98}.  For linewidths
of a few \kms, the magnetic hyperfine components within each of the
five features will blend together to form five electric
quadrupole hyperfine components consisting of a main line and two
symmetric pairs of satellite lines.  In this paper, we focus on large
intrinsic linewidths and therefore will only concern ourselves with
this ``five-peak'' profile.  These quadrupole satellite lines are
located from 10 to 30 \kms ~from the main line (see Table \ref{table}
for relative spacings and intensities).  Comparison of the intensity
of a satellite line to the main line enables a direct calculation of
the opacity of the gas making the NH\3 rotation inversion transitions
especially useful.  The derived opacity can then be used to calculate
$N_H$, masses, and also combined with line ratios of different
transitions to determine the rotational temperature of the gas
\citep{ho83}.

The small frequency separation of the five quadrupole hyperfine lines
means that for large intrinsic linewidths, $\dvi$, the satellite and
main hyperfine components will be blended into a single line.  Figure
\ref{prof.fig} shows line profiles for NH\3(1,1) for a range of
intrinsic linewidths for $\tm\ll1$.  For $\dvi\ga 5$~\kms ~the five
components are blended into a single profile and it becomes difficult
to measure the relative intensities needed for opacity estimation.
The situation is aggravated by the fact that an increase in opacity
will result in an increase in the intensity of the satellite hyperfine
lines relative to the main line and the blended profile will appear to
broaden even more.  As a result, an increase in either $\tm$ or $\dvi$
will appear as an increase in the observed linewidth, $\dvo$, and it
is difficult to interpret the data.  Exaggerated estimates of opacity
as a result of mis-fitting intrinsically broad lines can greatly limit
the physical parameters that can be determined from the data.  For
example, if NH\3(1,1) gas has $\dvi>5$~\kms, then an increase in
linewidth may be mistaken as an increase in opacity.  The
overestimation of the opacity propagates into overestimations of the
column density and ultimately the calculated mass of the gas.  In the
case of our Galactic Center data, the observed linewidths are high and
the challenge is to determine the extent of the contribution from
shocked material with large intrinsic linewidths as well as that from
high opacity gas.  Shocks are important for identifying the locations
of possible interactions between the streamers of molecular material
that surround the Galactic Center.  A mistaken calculation of the mass
of a cloud or incorrect identification of a shock could affect our
general understanding of the environment in the central parsecs of the
Galaxy.  It is therefore obvious that we need an improved method to
obtain independent estimates of the opacity and intrinsic linewidth of
the gas.

In this paper, we present a method to disentangle the effects of large
intrinsic linewidths and high opacities by comparing the observed
linewidths and intensities of two NH\3 transitions.  Because the exact
shape of the line profile need not be known, this method can be easily
applied to historical data where the line profiles are no longer
available.  Throughout the paper, we combine the theoretical
discussion with a test case example using our NH\3(1,1), (2,2), and
(3,3) data from the Galactic Center.  In \S \ref{corr}, we show the
correlation between $\dvo$ and $\tm$ in our Galactic Center data when
the two quantities are solved for in the traditional manner. Section
\ref{dvodvi} then presents a calculation of the dependence of $\dvo$
on $\dvi$ for NH\3(1,1), (2,2), and (3,3) for the special case where
$\tm\ll1$.  This is followed by the incorporation of variations in
opacity. The full method for simultaneously solving for the opacity
and intrinsic linewidth is presented in \S\ref{method}.  The paper
concludes by presenting the maps of the best estimate of $\dvi$ and
NH\3(1,1) opacity for our Galactic Center data.  The map of intrinsic
linewidths shows $\langle\dvi\rangle=15$ \kms ~for NH\3(1,1) which is
in agreement with observations of other tracers.  In addition, the
corrected opacities show that all except the most massive features in
the map are optically thin.  The opacity and intrinsic linewidths of
the observed features are then discussed.

\section{The Observed Correlation between $\dvo$ and $\tm$ \label{corr}}

Gas near the Galactic center has been observed to have intrinsic
linewidths of 10--20 \kms, and therefore provides a good test of the
effects of blending of hyperfine lines on the calculation of the
opacity \citep{gus80,arm85}.  We observed the central 10 pc of the
Galaxy in NH\3(1,1), (2,2), and (3,3) in 1999 March \citep{mcg01}.
The data were taken at 23~GHz in the D-northC configuration of
the Very Large Array (VLA) to provide the largest and most circular
beam possible.  A Gaussian taper was applied to the $uv$ data
resulting in a synthesized beam of $15''\times13''$, which is useful
for detection of extended features in our maps.  The data have a
velocity coverage of $\pm 150$ \kms, which includes almost all
detected motions at the Galactic Center, and a velocity resolution of
9.8~\kms.  In addition, a five pointing mosaic ensures full spatial
coverage of the region.  We choose the rotation inversion transitions
of NH\3 so that we can directly use the hyperfine lines to calculate
the opacity of the gas.  The opacity is then combined with line ratios
to calculate the rotational temperature of features observed in the
central 10 pc \citep{ho83}.  Increased temperatures can be a sign of
shocked gas which is important in determining if there are cloud-cloud
collisions or shocks from the expanding supernova remnant, Sgr A East
\citep{gen90,ho91,ser92,zyl99}.  Accurate measurements of the
intrinsic linewidth of the gas are also important for locating shocked
regions.  The large photon flux from the nucleus may also heat gas,
and features close to the Galactic Center should appear warmer.  All
of these quantities depend on accurate measurements of the opacity and
intrinsic linewidth of the gas.

We initially solved for the opacity in the traditional manner by
comparing the flux density of the main component to the average flux
density of the two outermost hyperfine components ($\pm19.3$ \kms ~for
NH\3(1,1)).  By averaging these two lines, we increase the S/N by a
factor of $\sqrt{2}$ and opacities are only calculated where this
average hyperfine flux density is greater than
$3/\sqrt{2}\sigma_{\rm{JK}}$ where $\sigma_{11}=0.009$ Jy Beam$^{-1}$,
$\sigma_{22}=0.009$ Jy Beam$^{-1}$, and $\sigma_{33}=0.011$ Jy
Beam$^{-1}$.  From radiative transfer, the opacity is related to the
observed flux density of the main and satellite hyperfine lines by

\bq
\frac{S(\nu)_m}{S(\nu)_s}=\frac{[\Delta T_A (\nu)]_m}{[\Delta T_A (\nu)]_s}=\frac{1-e^{-\tau(\nu)_m}}{1-e^{-a\tau(\nu)_m}}
\label{eq:tau}
\eq

\noindent where $S(\nu)_m$ and $S(\nu)_s$ refer to flux densities of
the main and satellite lines, $[\Delta T_A (\nu)]_m$ and $[\Delta T_A
(\nu)]_s$ are the respective antenna temperatures, $\tau(\nu)_m$ is
the opacity of the main hyperfine component, and $a$ is the
theoretical intensity of the satellite line compared to the main line
for $\tau(\nu)\ll1$.  Note that this equation assumes equal beam
filling factors and excitation temperatures for the two transitions.
The theoretical relative intensities of the quadrupole satellite
hyperfine lines for NH\3(1,1), (2,2), and (3,3) are listed in Table
\ref{table}.

Meanwhile, an observed linewidth, $\dvo$ (defined in this paper as the
FWHM), is fit to every pixel in the map by fitting a Gaussian of the
form \bq f(\rm{v})=A_0
e^{\frac{-(\rm{v}-\rm{v}_0)^2}{2\sigma_{obs}^2}} \eq
\noindent where $\rm{v}_0$ is the peak velocity at each position and
$\dvo=2.35\sigma_{obs}$.  Observed linewidths are only calculated for
pixels where the peak flux in the channel corresponding to $\rm{v}_0$
is greater than $3\sigma_{\rm{JK}}$.  The calculated observed
linewidths are also required to be less than 100 \kms.  The NH\3(1,1)
data give $\langle\dvo\rangle=30$~\kms ~which is approximately a
factor of two larger than what is observed in other tracers
\citep{gus80,arm85,gen87}.  Linewidths calculated from the NH\3(3,3)
data, however, have $\langle\dvo\rangle=18$~\kms.  For $\tm\ll1$, the
NH\3(1,1) satellite hyperfine lines are a much larger fraction of the
main line intensity (28 and 22\%) than in the case of NH\3(3,3)
($\sim3$\%).  In addition, the NH\3(1,1) quadrupole hyperfine lines
have a much smaller velocity separation than the higher transitions
(see Table \ref{table}).  These two characteristics of NH\3(1,1) make
it much more susceptible to corruption by blending of the quadrupole
hyperfine lines at large intrinsic linewidths.  The difference in the
observed linewidths for the NH\3(1,1) and (3,3) data may thus be
explained by the large intrinsic linewidths near the Galactic Center.

\begin{figure}
\plottwo{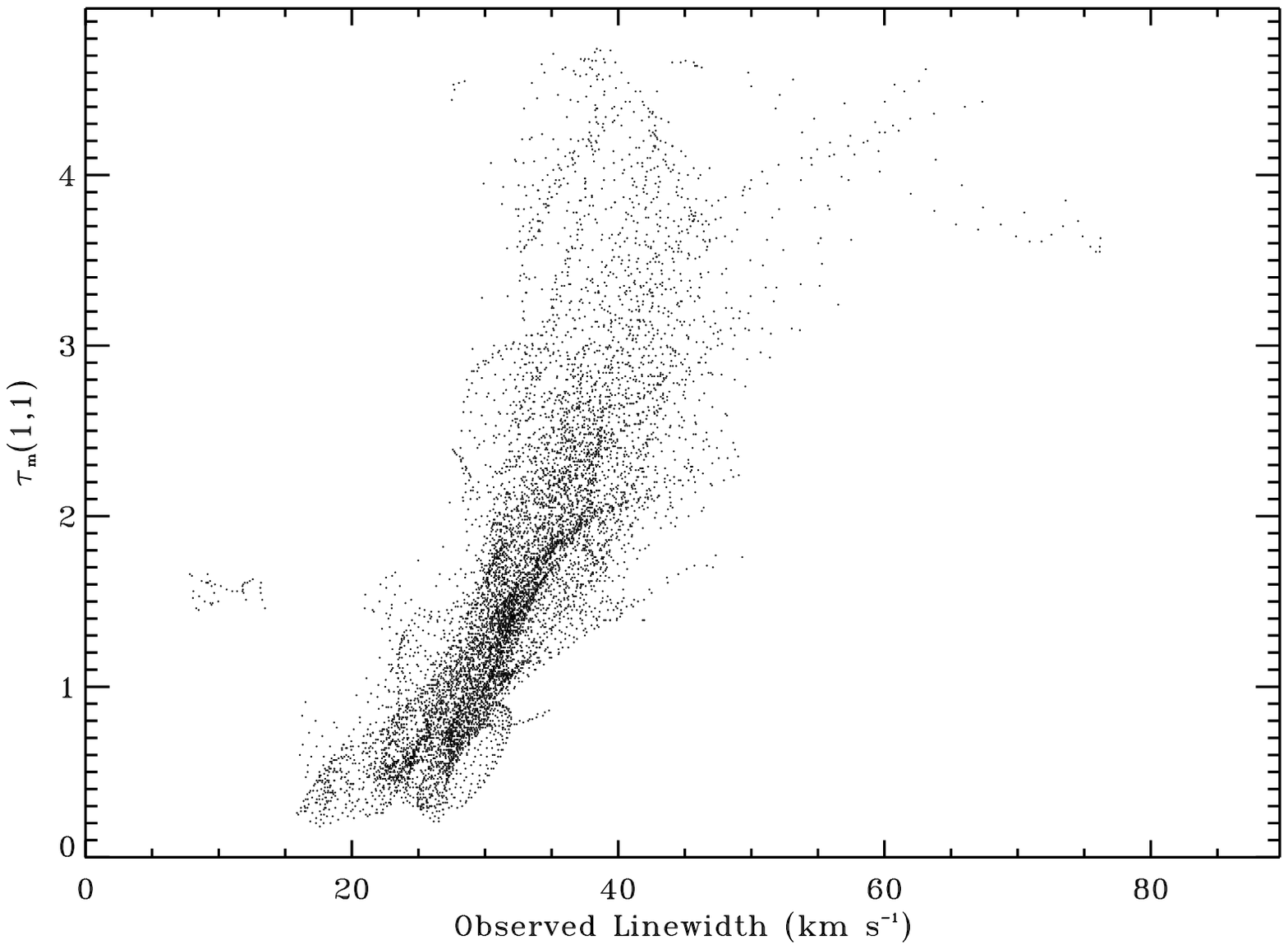}{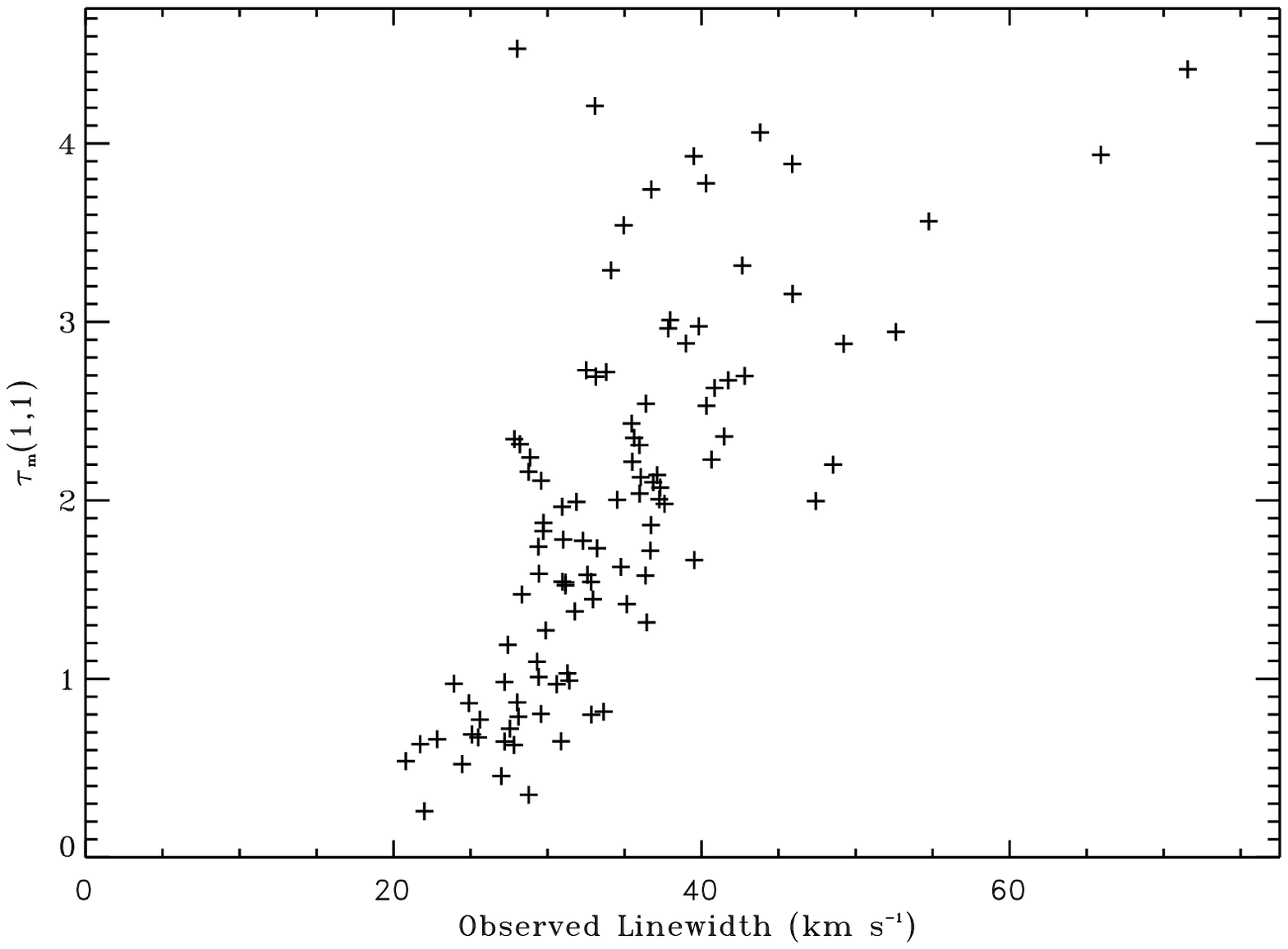}
\caption{$a)$ Observed NH\3(1,1) linewidth, $\dvoa$, as a function of
$\ta$ for all points in our data where the flux was bright enough to
fit for both parameters.  The strong correlation indicates that
blending due to large linewidths has corrupted our results such that
an increase in observed linewidth also appears as an increase in
opacity.  The correlation between adjacent points in the plot is the
result of many samplings per beam ($\theta_b=15''$). $b)$ Same as $a$,
but the data were first averaged over a single beam and the resulting
values for $\ta$ and $\dvo$ are plotted. The correlation between $\ta$
and $\dvo$ is still easily observable and has a correlation
coefficient of 0.69. \label{corr.fig}}
\end{figure}

\begin{figure}
\plotone{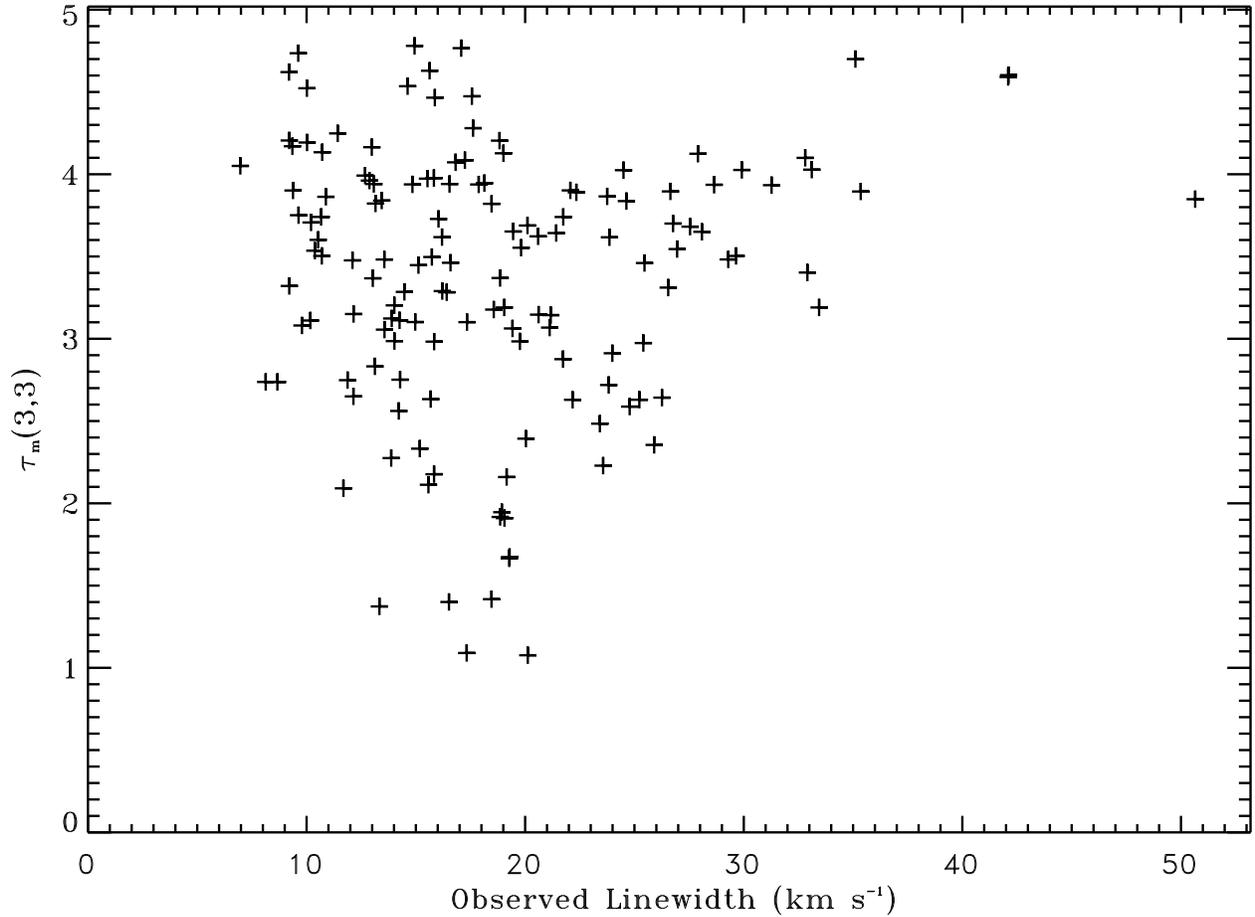}
\caption{Observed NH\3(3,3) linewidth, $\dvoc$, as a
function of $\tc$ for all points in our data where the flux was bright
enough to fit for both parameters.  The correlation is less strong
than in Figure \ref{corr.fig} and has a correlation coefficient of
0.07. \label{33corr.fig}}
\end{figure}

To test whether blending has affected the NH\3(1,1) data, we plot
$\dvo$ as a function of the main line opacity, $\ta$ for every pixel
in our NH\3(1,1) map where $\dvo<100$~\kms ~and $\ta<5$ (see Figure
\ref{corr.fig}$a$).  The main line opacity has been calculated using
Equation \ref{eq:tau} and the relative intensities of the main and
outer satellite hyperfine lines.  The noise in one channel,
$\sigma_{11}$, results in an estimated error of $\sim0.1$ for $\ta$.
The error in $\dvo$ is 1--2~\kms (see \S\ref{errors}).  A strong
correlation between the two quantities is observed.  However, there
are many pixels per $\sim15''$ beam resulting in correlation between
adjacent pixels.  In Figure \ref{corr.fig}$b$, we average over the
beam and sample once per beam resulting in independent pixels.  The
correlation coefficient between $\dvo$ and $\ta$ is 0.69.  Figure
\ref{33corr.fig} shows the same parameters for NH\3(3,3) and the
correlation coefficient between $\dvo$ and $\tc$ is only 0.07.  The
fact that the correlation is weaker in NH\3(3,3) than (1,1) suggests
that the correlation between $\dvo$ and $\ta$ is not real, but is an
artifact of blending of the hyperfine components.  It is interesting
to note that there are no points with $\tc<1.0$ in this figure.  This
effect is explained by the minimum flux cutoff of $3\sigma_{33}$ (0.03
Jy beam$^{-1}$) required for the average outer satellite hyperfine
line.  Because the satellite hyperfine lines are intrinsically weaker
in NH\3(3,3) than in (1,1) (see Table \ref{table}), the (3,3)
hyperfine lines are often below this noise cutoff resulting in an
effective cutoff in the minimum $\tc$ to which we are sensitive.


\section{Dependence of $\dvo$ on $\dvi$ and $\tm$} 
In order to develop a method to remove the effects of blending on
$\dvo$, we must model the dependence of the observed linewidth on both
the intrinsic linewidth and the opacity of the gas.  The opacity of
the gas as a function of velocity is defined as

\bq 
\tau(\rm{v})=\tau_m \phi(\rm{v},\sigma) 
\label{tau.eq}
\eq

\noindent where $\tm$ is the opacity at the center of the main line
and $\phi(\rm{v},\sigma)$ is the normalized line profile for the
rotation inversion transition.  The quadrupole hyperfine components
are modeled by five Gaussians with velocity spacings and relative
intensities of the satellite hyperfine lines defined as in Table
\ref{table}.  The profile is written explicitly as

\begin{eqnarray}
\phi(\rm{v},\sigma)=&\,&e^{-\rm{v}^2/2\sigma_{int}^2} \nonumber\\
&+&a_{\rm{in}}e^{-(\rm{v}-\Delta\rm{v}_{in})^2/2\sigma_{int}^2}+a_{\rm{in}}e^{-(\rm{v}+\Delta\rm{v}_{in})^2/2\sigma_{int}^2} \nonumber\\
&+&a_{\rm{out}}e^{-(\rm{v}-\Delta\rm{v}_{out})^2/2\sigma_{int}^2}+a_{\rm{out}}e^{-(\rm{v}+\Delta\rm{v}_{out})^2/2\sigma_{int}^2}
\label{phi.eq}
\end{eqnarray}

\noindent where $\sigma_{\rm{int}}=\dvi/2.35$ is the standard
deviation of the line profile, $\Delta \rm{v}_{in}$ and $\Delta
\rm{v}_{out}$ are the offsets of the inner and outer satellite
hyperfine lines (in \kms) respectively, and $a_{\rm{in}}$ and
$a_{\rm{out}}$ are the intensities of the inner and outer satellite
hyperfine lines assuming an intensity of 1.0 for the main line (see
Table \ref{table}).  The intensity of the line is given by

\bq 
\inten=(T_{ex}-T_b)[1-e^{-\tau_m\phi(\rm{v},\sigma)}]~.
\label{int.eq}  
\eq

Throughout the paper, when the opacity is referred to in the general
case, we use $\tm$.  For a specific (J,K) transition of NH\3, the main
line opacity is written as $\tm$(J,K).  It is this opacity of the main
line, $\tau_m$, that is reported in most observational papers and in
keeping with this convention, the opacities derived in this paper all
refer to $\tau_m$.

\subsection{Dependence of $\dvo$ on $\dvi$ for $\tm\ll1$ \label{dvodvi}}

We begin by calculating the line profile as a function of $\dvi$ in
the special case where $\tm\ll1$.  We need not be concerned with the
observed value of $\inten$ because the FWHM and relative intensity of
the main and satellite hyperfine lines are independent of the absolute
amplitude of the line.  Therefore, Equation \ref{int.eq} reduces to
$\inten\propto\phi(\rm{v},\sigma)$ for $\tm\ll1$.  Figure
\ref{prof.fig} compares the line profiles of NH\3(1,1) for $\tm\ll1$
for a range of intrinsic linewidths.  For $\dvi\ga5$ \kms, the main
and satellite components of NH\3(1,1) blend into a single line
profile and fitting a Gaussian profile to the main component will
overestimate the intrinsic linewidth of the gas.

\begin{figure}
\plottwo{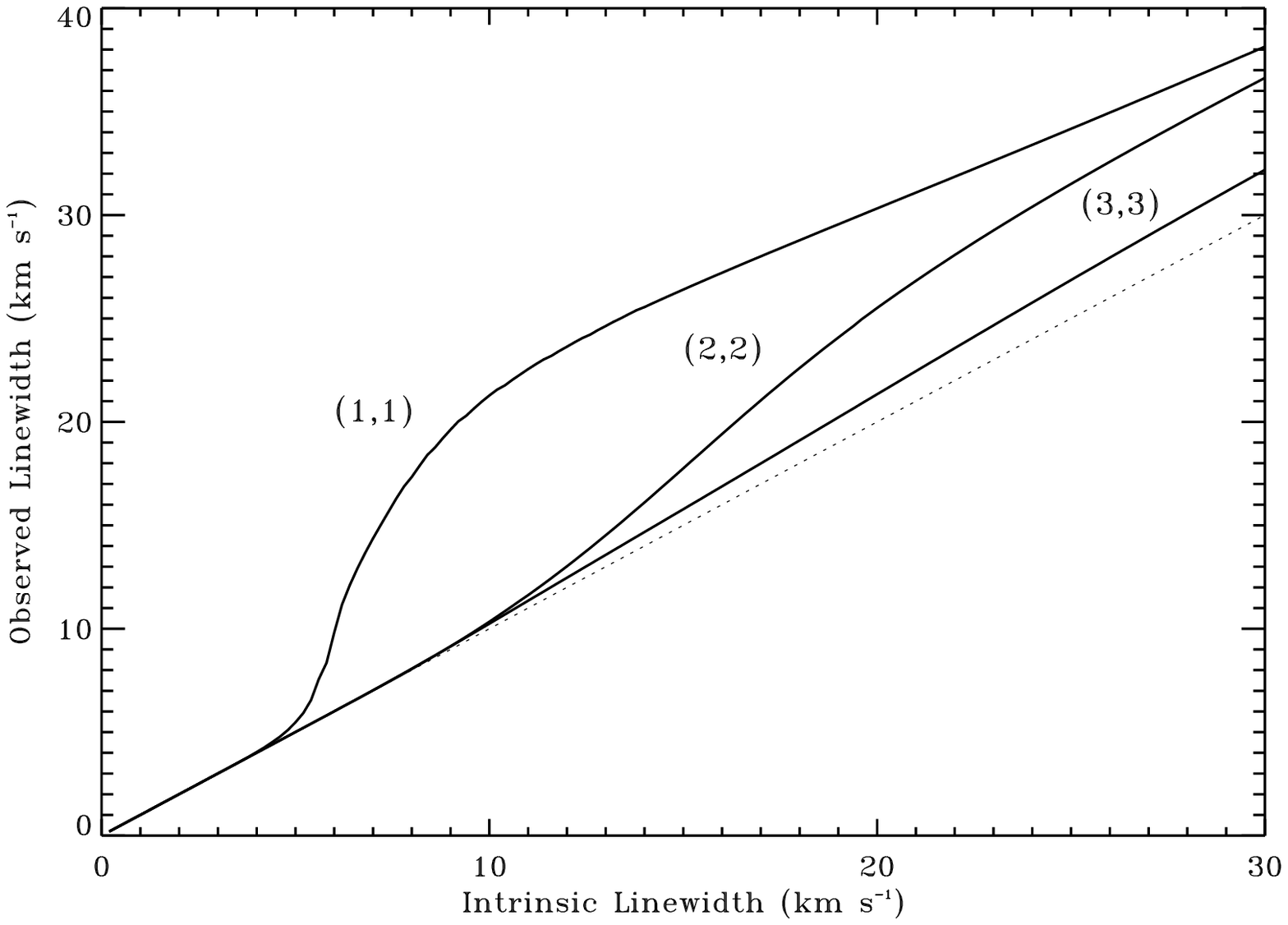}{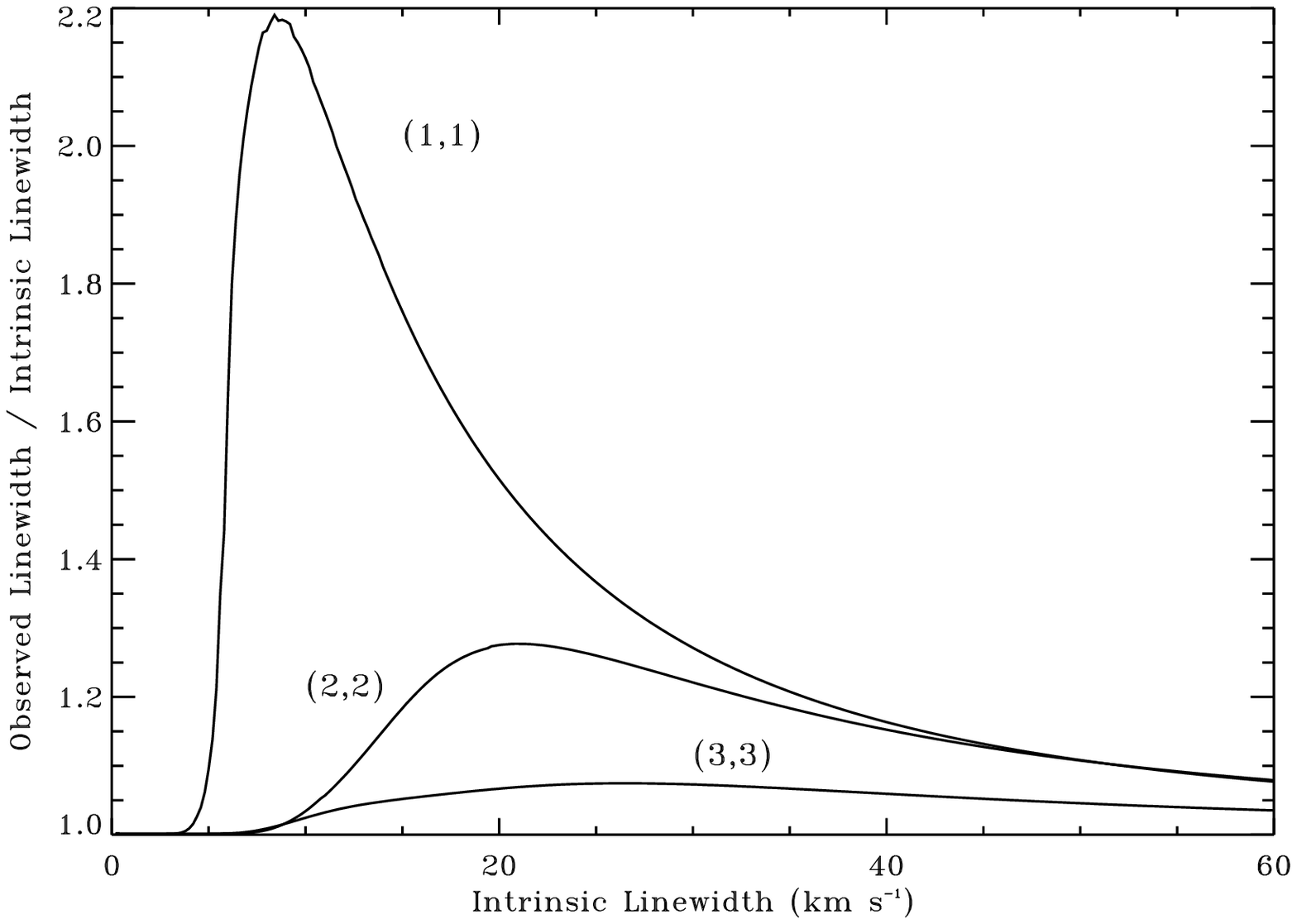}
\caption{$a)$Observed linewidth as a function of intrinsic linewidth
for NH\3(1,1), (2,2), and (3,3) for $\tm\ll1$. The dashed line shows
$\dvo=\dvi$.  Note that the observed linewidth of NH\3(1,1) will
greatly overestimate $\dvi$ for $\dvi>5$~\kms. $b)$ $\dvo/\dvi$ for
$\dvi\le60$~\kms ~for NH\3(1,1), (2,2), and (3,3) for
$\tm\ll1$. \label{dvodvi.fig}}
\end{figure}

In order to determine $\dvo$ as a function of $\dvi$ for NH\3(1,1),
(2,2), and (3,3), normalized line profiles as a function of $\dvi$ and
v are generated for the case where $\tm\ll1$.  For each normalized
line profile, a Gaussian is fit to the main component to determine the
resulting observed linewidth.  By modeling the effect of $\dvi$ and
$\tm$ on the observed linewidth, the method for recovering the
intrinsic linewidth can be applied to any historical data where the
entire line profile may no longer be available.  Figure
\ref{dvodvi.fig}$a$ shows $\dvo$ as a function of $\dvi$ for NH\3
(1,1), (2,2), and (3,3) assuming $\tm\ll1$.  For $\dvi\la5$ \kms, no
blending occurs and all three transitions have $\dvo\approx\dvi$.
Note that we are not concerned with the case of dark cores where
intrinsic linewidths of only a few \kms ~cause the closely separated
18 magnetic hyperfine lines to blend, which is discussed in detail in
\citet{ho77} and \citet{bar98}.

For intrinsic linewidths greater than 5 \kms, the satellite hyperfine
lines of NH\3(1,1) blend with the main line and $\dvoa$ will greatly
exaggerate $\dvi$ for $\dvi\ga5$ \kms, {\it independent of opacity and
velocity resolution.}  NH\3(2,2) and (3,3) give a much better
estimation of the intrinsic linewidth for $\tm\ll1$ because the
satellite hyperfine lines are relatively weaker compared to the main
lines and are also located further from the main line than in the case
of NH\3(1,1).  In our Galactic Center data, we find
$\langle\dvo\rangle=30$~\kms ~for NH\3(1,1) and
$\langle\dvo\rangle=18$~\kms ~for NH\3(3,3).  These results can be fit
almost exactly by $\dvi=17$~\kms ~in Figure \ref{dvodvi.fig}$a$.

Figure \ref{dvodvi.fig}$b$ plots $\dvo/\dvi$ for intrinsic linewidths
up to 60~\kms.  For each transition, there exists a velocity,
$\rm{v_{crit}}$, above which the exaggeration of $\dvi$ by $\dvo$
decreases.  As the intrinsic linewidths approach $\rm{v_{crit}}$, the
line profile is broadening as the satellite hyperfine lines blend with
the main line.  However, once the intrinsic linewidth of the gas is
greater than the maximum separation of the hyperfine lines, the effect
of the satellite hyperfine lines on $\dvo$ becomes less important and
$\dvo$ asymptotically approaches $\dvi$.  The critical velocities for
NH\3(1,1), (2,2), and (3,3) are 8, 20, and 27~\kms, respectively.
For the Galactic Center where $\dvi\approx15$~\kms, NH\3(1,1) gives
by far the worst overestimation of $\dvi$.  However, for
$\dvi\simgt50$~\kms, NH\3(2,2) is equally as bad as NH\3(1,1). For
the remainder of this paper, we focus on $\dvi\le30$~\kms, as expected
at the Galactic Center.

\subsection{Observed line profiles as a function of $\dvi$ and $\tau$ \label{prof}}

\begin{figure}
\plotone{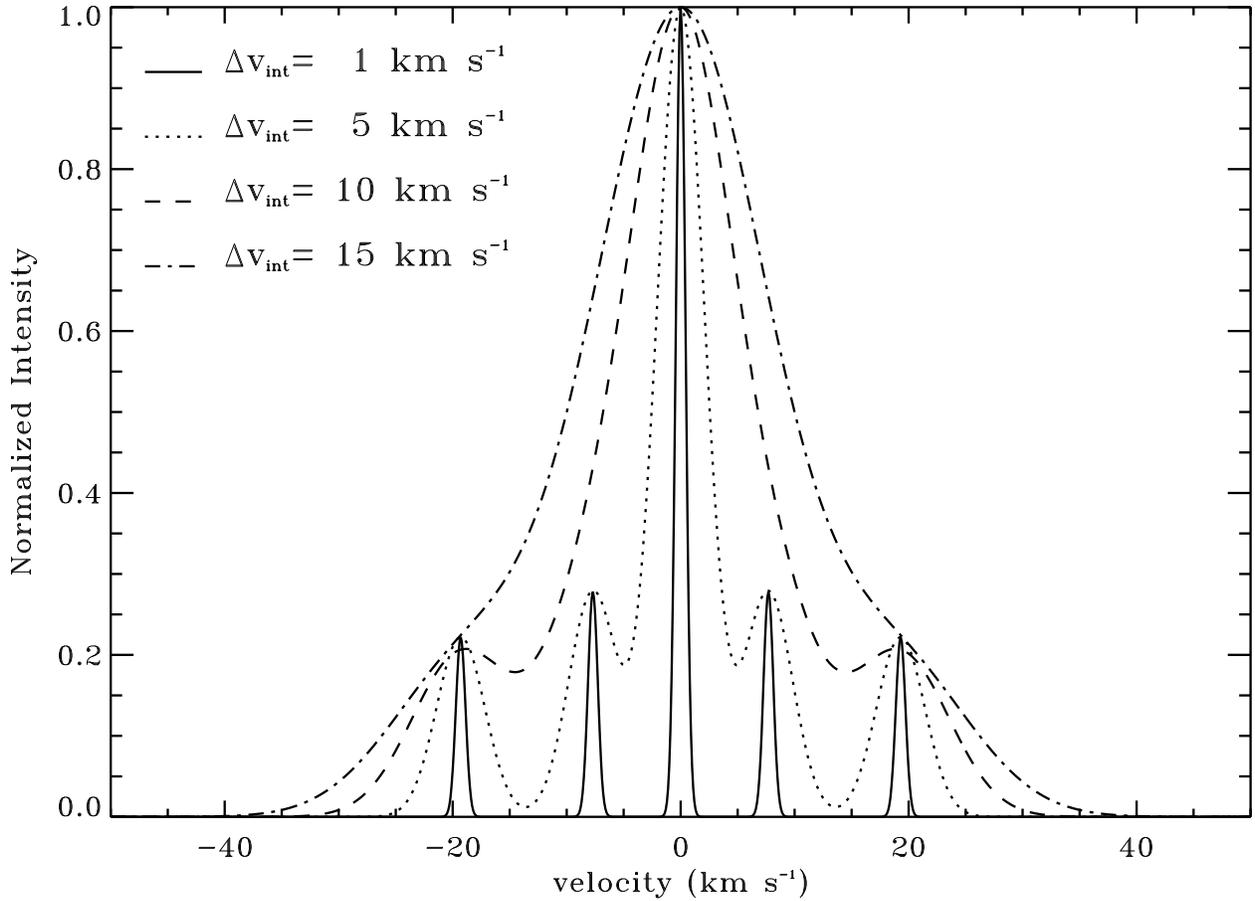}
\caption{Line profiles for NH\3(1,1) for $\tm\ll1$ and
$\dvi=1$, 5, 10, and 15~\kms.  As $\dvi$ increases, the five electric
quadrupole hyperfine components blend into one single
line. \label{prof.fig}}
\end{figure}

\begin{figure}
\plottwo{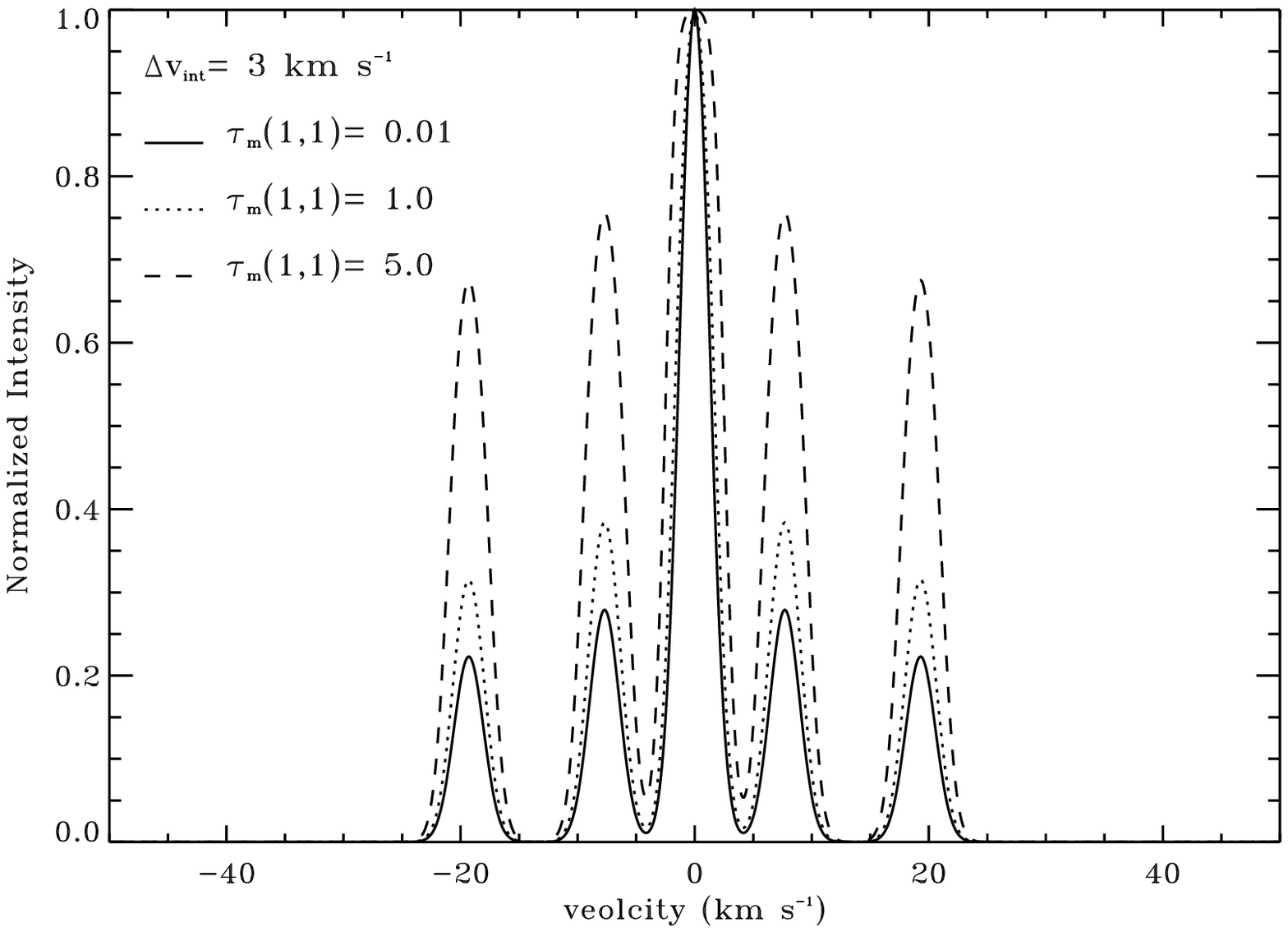}{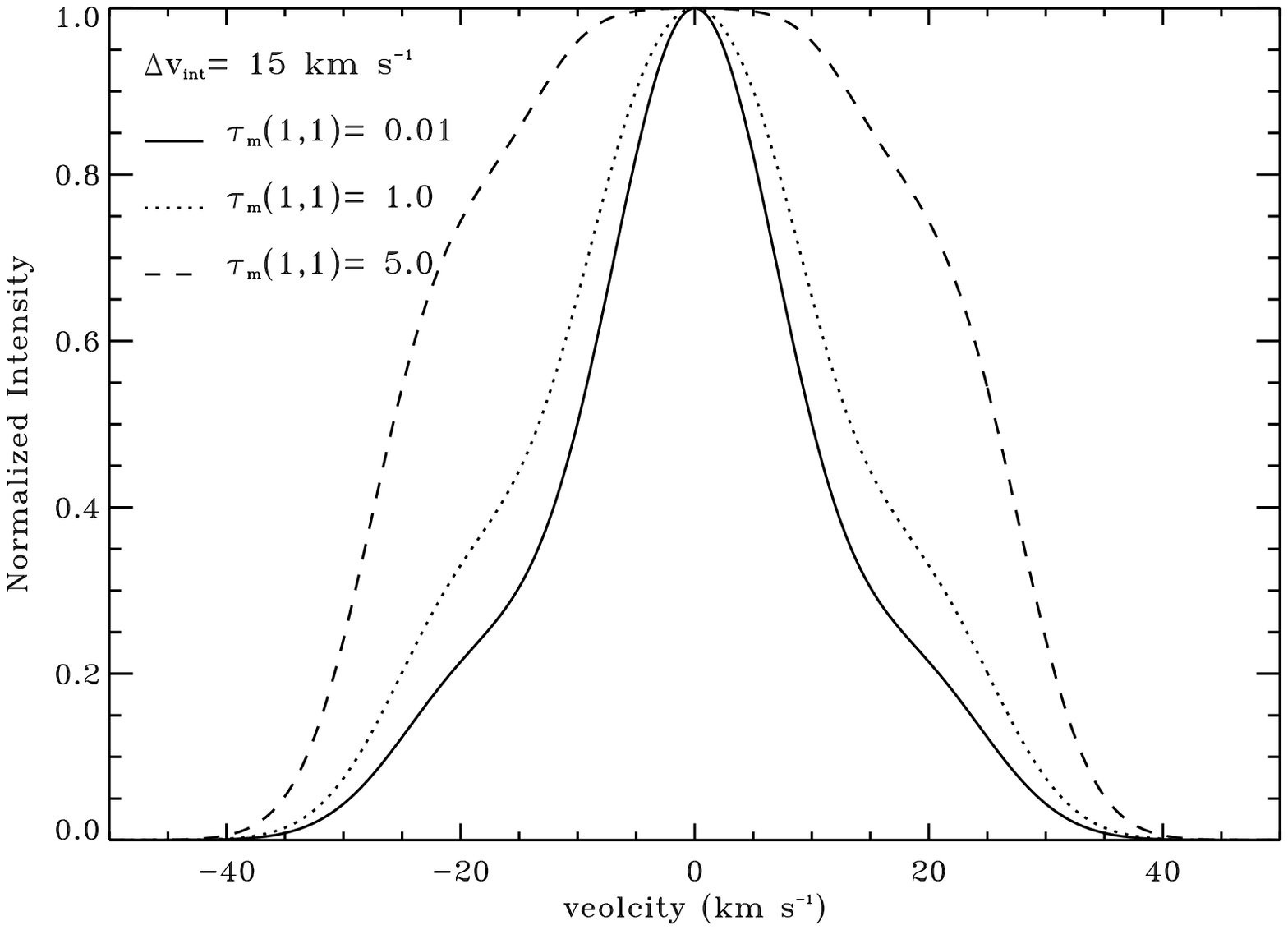}
\caption{$a)$ Line profiles of NH\3(1,1) for
$\tau=0.01$, 1.0 and 5.0 for $\dvi=3$ \kms.  In this case, there is
no blending so a fit of a Gaussian to the main line will give a good
estimate of $\dvi$ (see Figure \ref{123dvo.fig}$a$).  $b)$ Same as $a$,
but for $\dvi=15$~\kms. In this case, the hyperfine components are
highly blended and an increase in $\ta$ will greatly affect the
observed linewidth.  \label{dtau.fig}}
\end{figure}

In this section, we calculate line profiles for the case where the
opacity is large enough that the intensity of the satellite hyperfine
lines becomes important.  In the case of the blended profile, an
increase in opacity will appear as an increase in the observed
linewidth of the profile.  Figures \ref{prof.fig} and \ref{dtau.fig}
can be used to compare the effects of changes in either $\dvi$ or
$\tau_m$ while the other parameter is held constant.  As seen in
Figure \ref{prof.fig}, an increase in $\dvi$ for constant opacity
results in broadening of the line profile until the five quadrupole
hyperfine lines blend together.  Figures \ref{dtau.fig}$a$ and
\ref{dtau.fig}$b$ show the effect of varying $\tau_m$ when $\dvi=3$
and 15~\kms ~for NH\3(1,1).  In Figure \ref{dtau.fig}$a$,
$\dvi<5$~\kms ~so the profiles are not blended.  As a result, an
increase in $\tau_m$ only results in an increase in the relative
intensities of the satellite lines with respect to the main line.  In
Figure \ref{dtau.fig}$b$, however, the blended profile appears to
broaden dramatically as the opacity is increased.

\begin{figure}
\plotone{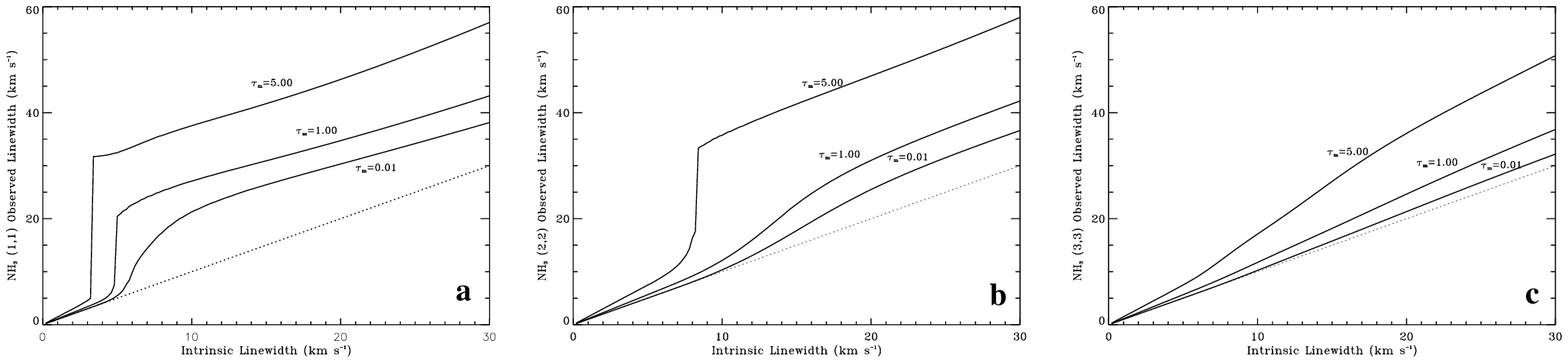}
\caption{$\dvo$ as a function of $\dvi$ for $\tm=0.01, 1.0, 5.0$ for
$a)$ NH\3(1,1), $b)$ NH\3(2,2) and $c)$ NH\3(3,3). The dashed line
shows $\dvo=\dvi$.  For large $\tm$, a discontinuity develops at
$\sim3$~\kms ~for NH\3(1,1) and $\sim8$~\kms ~for NH\3(2,2) where
blending of hyperfine lines with comparable intensities suddenly
results in a very broad line profile. \label{123dvo.fig}}
\end{figure}

Using the definition of the line profile in Equation \ref{int.eq}, an
array of line profiles is generated for a range of opacities and
intrinsic linewidths.  If the data have good velocity resolution with
many channels across the profile, then the intrinsic linewidth and
$\tm$ of any NH\3 rotation inversion transition may be solved for
without any additional information from other tracers.  In this case,
a least squares minimization routine can be used to search this array
to find the $\dvi$ and $\tm$ that give the best fit to the observed
line profile.  In the case where multiple transitions of NH\3 are
observed, similar results may be easily obtained by comparing only the
observed linewidths and intensities of the lines.  For each line
profile in the generated data, the main line is fit with a Gaussian to
estimate the expected value for $\dvo$.  Figure \ref{123dvo.fig} shows
the dependence of $\dvo$ on $\dvi$ and $\tm$ for the simulated NH\3
(1,1), (2,2), and (3,3) data. For small intrinsic linewidths, the
hyperfine components are well-separated and $\dvo\approx\dvi$,
independent of $\tm$.  A discontinuity develops near 5~\kms ~for NH\3
(1,1) and 8~\kms ~for NH\3(2,2) in the optically thick cases
resulting from the blending of main and satellite hyperfine components
with comparable intensities (see the line profile for $\ta=5.0$ in
Figure \ref{dtau.fig}$a$).  NH\3(3,3) does not show this effect in
our models because the satellite hyperfine lines are $<20\%$ of the
intensity of the main component even in the case where $\tc=5.0$.
Using Figure \ref{123dvo.fig}$a$, we find that for an average
intrinsic linewidth at the Galactic Center data of 17~\kms ~as in \S
\ref{dvodvi}, the average observed linewidth of 30~\kms ~in our data
indicates that $0.01<\ta<1.0$.

\section{Method to Simultaneously Determine $\dvi$, $\ta$, and $\tc$ \label{method}}

In this section, we outline the detailed steps of our method for
recovering the intrinsic linewidths and opacities of gas with large
linewidths using our Galactic Center data as an example.  We use NH\3
(1,1) and (3,3) because they are the brightest tracers in our data and
thus have the most points for which $\dvo$ can be calculated for both
transitions.  There is much NH\3(1,1) emission because it traces the
lowest energy of the rotation inversion transitions (23.4 K above
ground) \citep{ho83}. The NH\3(3,3) image is bright, however, due to
a quantum degeneracy in the `ortho' ($J=3\rm{n}$) form of NH\3 which
increases its intensity by a factor of two \citep{ho83}.  In addition,
high temperatures near the Galactic Center make this transition
well-populated.

Two assumptions must hold for the method to be valid.  First, it is
assumed that the two transitions have the same intrinsic linewidth.
Our NH\3(1,1) and (3,3) velocity integrated maps display a very high
degree of correspondence and it seems valid to postulate that the two
transitions trace the same material and therefore should have the same
intrinsic linewidth.  In addition, it is necessary to assume
that the opacity does not exceed a particular value, $\tau_{max}$, in
either transition.  This maximum opacity is used to limit the size of
the parameter space which must be searched.  In the case below,
we have assumed $\tau_{max}=5$ for both NH\3(1,1) and (3,3).  As seen
in \S \ref{gc}, almost all points in our dataset have final opacities
less than 1.0, which is well below $\tau_{max}$.

The following four steps can be used to recover the intrinsic
linewidth and opacity of NH\3 rotation inversion transitions when
observed linewidths and intensities of two transitions are known.

\noindent {\it (1)} A Gaussian profile is fit to the upper transition,
to obtain $\dvo$.  This upper transition is chosen because it will be
less affected by blending and will give the best initial estimate of
$\dvi$ (see Figure \ref{dvodvi.fig}$b$).  Using the NH\3(3,3) observed
linewidth, $\dvoc$, the relation between $\dvo$ and $\dvi$ can be used
to solve for a range of valid $\dvi$ corresponding to the case where
$0<\tc<\tau_{max}$ (see Figure \ref{123dvo.fig}$c$).  The range of
valid intrinsic linewidths is denoted here as

\bq
\dvilow<\dvi<\dvihigh~.
\eq

\noindent For example, Figure \ref{123dvo.fig}$c$ shows that if
$\dvoc=15$~\kms ~then $9$\kms$\la\dvi\la14$\kms ~for $0\le\tc\le5$.

\begin{figure}
\plotone{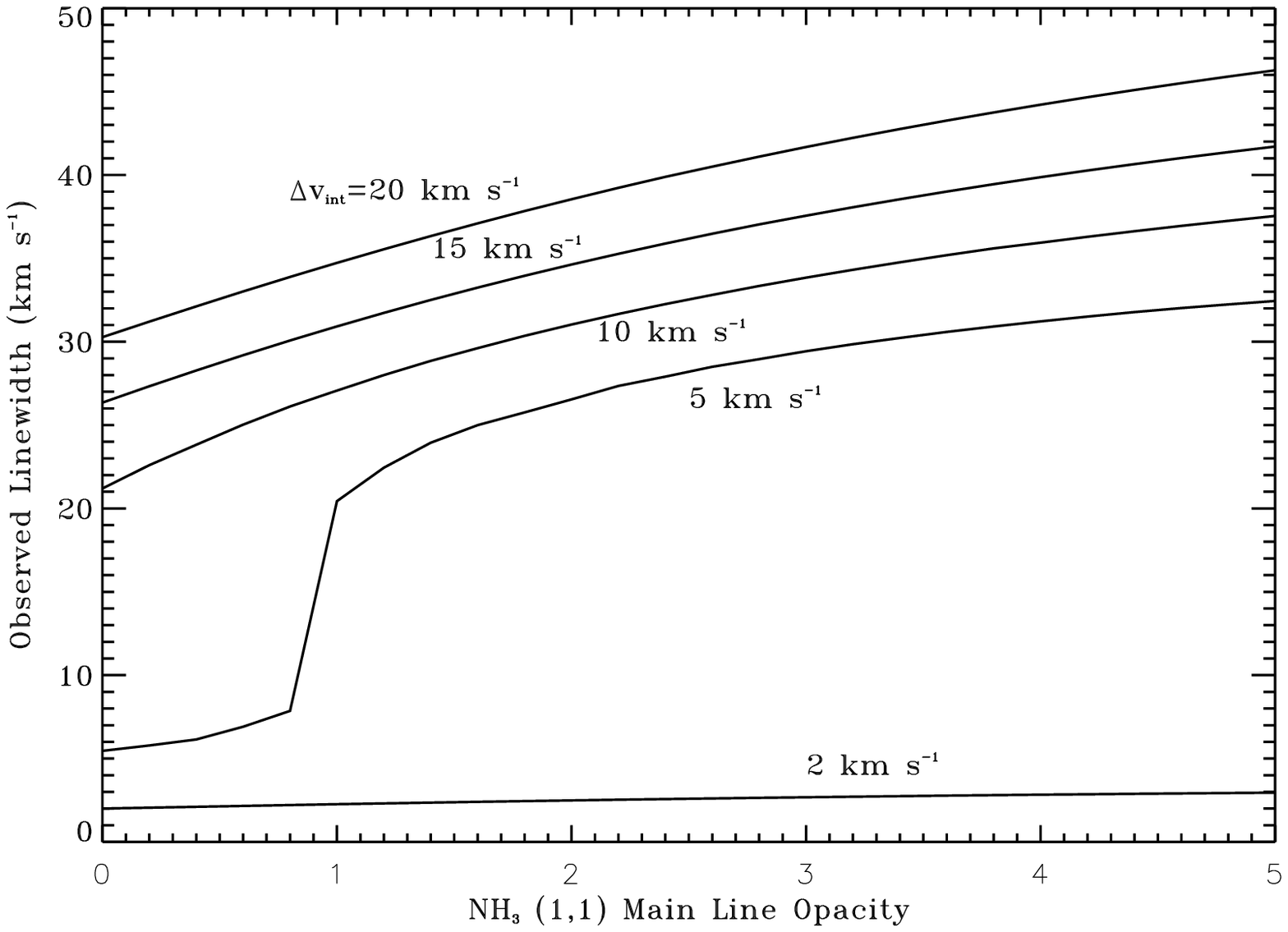}
\caption{Observed NH\3(1,1) linewidth as a function of $\ta$ for a
range of $\dvi$. \label{dvotau.fig}}
\end{figure}

\noindent {\it (2)} With a range of valid intrinsic linewidths, the
NH\3(1,1) observed linewidth, $\dvoa$, is used to find the range of
acceptable NH\3(1,1) opacities.  Each curve in Figure
\ref{dvotau.fig} corresponds to a different intrinsic linewidth.
Therefore Step 1 enables the determination of a range of valid curves
where $\dvilow<\dvi<\dvihigh$.  There is only a limited range of $\ta$
where the NH\3(1,1) observed linewidth lies between the minimum and
maximum curves in Figure \ref{dvotau.fig}.  In this way, $\dvoa$ is
combined with the range of intrinsic linewidths from Step 1 to
calculate a valid range for $\ta$ such that

\bq
\talow<\ta<\tahigh~.
\eq

\noindent In the example discussed above, we know
$9$\kms$\la\dvi\la14$\kms.  Now, if $\dvoa=30$~\kms ~then we can use
Figure \ref{dvotau.fig} to determine that $\ta$ must lie between 0.9
and 1.9.

\noindent {\it (3)} The NH\3(3,3) opacity can be calculated from the
NH\3(1,1) opacity and the ratio of the brightness of the main lines
of NH\3(3,3) to (1,1) using

\bq
\tm(J,K)=-\ln[1-\frac{[\Delta T_A (\nu)]_m(J,K)}{[\Delta T_A (\nu)]_m(1,1)}(1-e^{-\ta})]~.
\label{eq8}
\eq

\noindent As stated in \S \ref{corr}, this equation assumes equal beam
filling factors and excitation temperatures for NH\3(1,1) and (3,3).
Given the similarity of the velocity integrated maps for these two
transitions (see \citet{mcg01}), the assumption of equal beam filling
factors is reasonable.  In addition, the densities of
$10^4-10^5$~cm$^{-3}$ in these clouds should result in thermalization
that will tend to make the excitation temperatures of NH\3(1,1) and
(3,3) equal.  With these assumptions, a range of acceptable $\tc$ can
be directly calculated using Equation \ref{eq8} and the range of $\ta$
found in Step 2 such that

\bq
\tclow<\tc<\tchigh~.
\eq

\noindent These new limits on $\tc$ provide a tighter constraint than
our initial assumption that $0<\tc<\tau_{max}$.

\noindent {\it(4)} With a smaller range for $\tc$, we can return to Step
$1$ and determine a tighter constraint for $\dvi$.  It follows that
this process can be repeated until it converges to a solution for
$\dvi$, $\tc$, and $\ta$ within some predefined tolerance.  In our
data, we find that the solutions will converge in less than five
iterations for a tolerance of 0.4 for $\ta$.  The fast conversion of
this method makes it especially useful for datasets with many pixels.

\subsection{Estimated Errors}
\label{errors}

To estimate the error in $\dvi$ and $\tm$, we first estimate the error
in the fit of $\dvo$ to a noisy line profile.  Starting with the
theoretical line profile, $\rm{I}(\rm{v},\sigma,\tm)$, noise is added
at the 10\% level and the observed linewidth, $\dvo'$, is measured.
This observed linewidth is compared to the expected FWHM, $\dvo$,
found by fitting a Gaussian to the noiseless profile.  The process is
repeated until the spread in the differences between $\dvo'$
and $\dvo$ can be used to estimate the error in observed linewidth.  We
first model the errors in NH\3(1,1) in the case where the channel
width is much less than the separation of the hyperfine components.
The largest errors (up to 3~\kms) occur for $\dvi\approx5$~\kms ~and
$0.5<\ta<1.0$.  In this case, the five hyperfine components are
partially blended, making it difficult to properly estimate a
linewidth for the main line.  Elsewhere, the expected error in $\dvo$
is less than 0.5~\kms.  For $\dvi=15$~\kms, the error in the observed
linewidth is 0.3~\kms ~for a spectrum with $S/N=10$.  For NH\3(3,3),
the hyperfine components have a much smaller effect and the error in
the fit to $\dvoc$ is never greater than 0.5~\kms, independent of
$\dvi$ and $\tc$.

For the Galactic center data discussed below, the channel width of
9.8~\kms ~is close to the velocity separation of the hyperfine
components and dominates the error for $\dvi<5$~\kms.  However, our
observations were planed with the knowledge that intrinsic linewidths
at the Galactic center range between 10 and 20~\kms.  For
$\dvi\simgt5$~\kms, the fit to $\dvo$ is reasonable and the error
ranges between 1 and 2~\kms ~with a slight dependence on $\ta$ and
$\dvi$.  Assuming $\dvi=15$~\kms ~for the Galactic Center, the
$1\sigma$ error in the fit to $\dvoa$ and $\dvoc$ is 2~\kms and 1~\kms,
respectively.

Simulations of the error in $\dvi$ and $\ta$ were made using an error
of 2~\kms ~for $\dvoa$ and $\dvoc$ and an error of 0.01~Jy beam$^{-1}$ for
the amplitude of the NH\3(1,1) and (3,3) main lines.  We find a
typical error of 0.4 for $\ta$ and 4~\kms ~for $\dvi$ with the caveat
that the opacities and linewidths cannot be negative.  In both cases,
errors show a slight increase towards larger values of $\dvi$ and
$\ta$.  This results from the fact that the slope in Figures
\ref{123dvo.fig} and \ref{dvotau.fig} becomes more shallow for large
values of $\dvi$ and $\ta$.

\section{Results for Galactic Center Data}
\label{gc}

We have applied the method above to our NH\3(1,1) and (3,3) Galactic
Center data for all pixels with calculated linewidths less than 100
\kms.  Each pixel was calculated individually and then the resulting
map was smoothed with an eight pixel median filter to match the
resolution intrinsic to our data.  Because pixels for which no value
can be determined are set to zero, a convolution by the beam would
result in a halo of low values around all features in the map.  As a
result, every cloud would appear to be optically thin and have low
intrinsic linewidths at the edge which could be misleading.  The
median filter will not produce this halo and is therefore better
suited to be used in smoothing the data.

\begin{figure}
\plotone{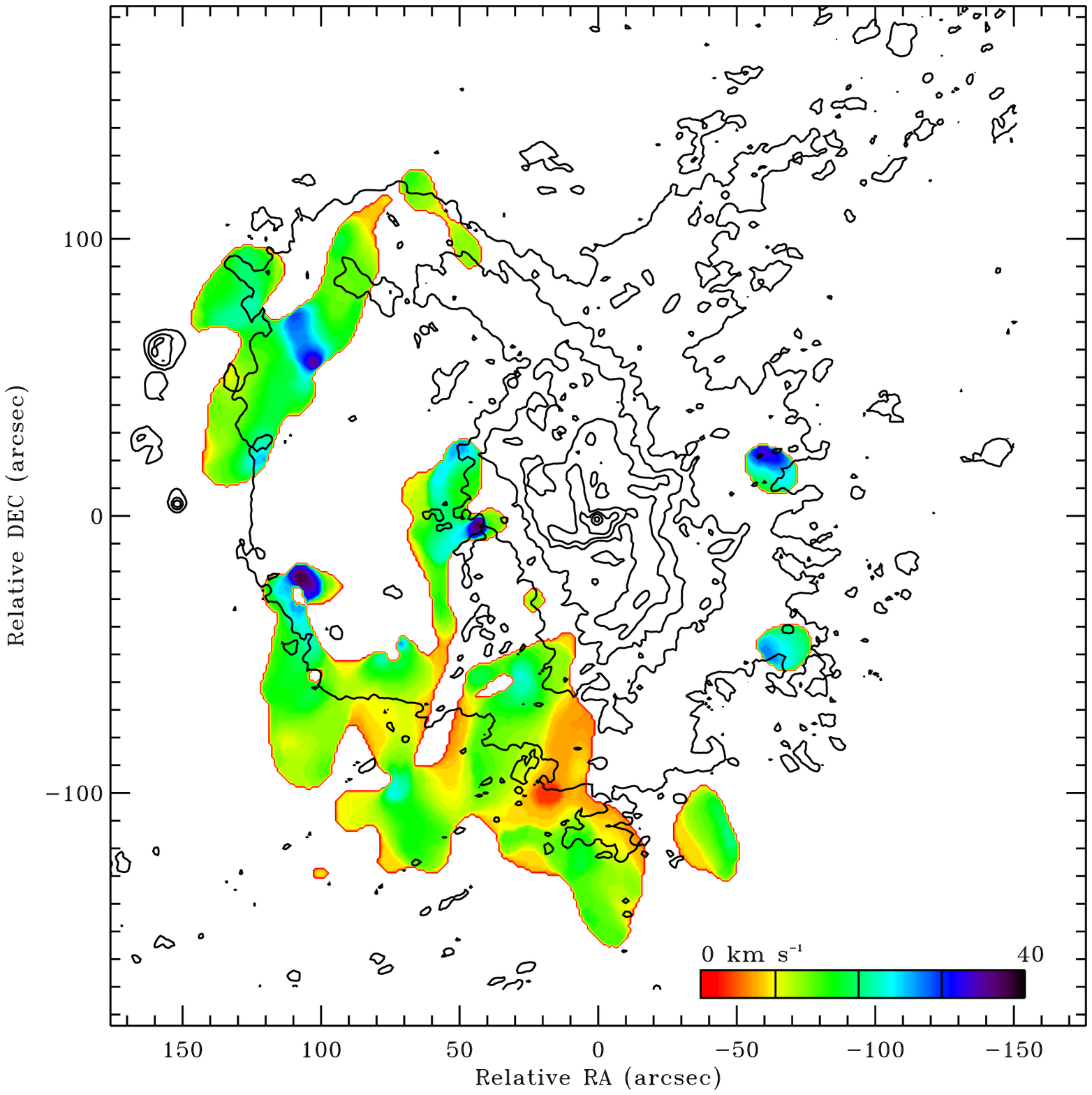}
\caption{Best estimate of $\dvi$ in color overlaid on an
image of 6~cm continuum emission in contours \citep{yus87}. Contours
are 0.001, 0.01, 0.02, 0.04, 0.08, 0.16, 0.32, 0.64, and 1.28 Jy beam$^{-1}$.
The linewidth map has been smoothed with an eight pixel median filter
(see \S \ref{gc}).  The continuum image shows the location of the
expanding shell, Sgr A East, the mini-spiral, and Sgr A*.  Typical
intrinsic linewidths are 15~\kms.  The largest $\dvi$ appear to be
located near the nucleus or on the edge of Sgr A East.
\label{dvi.fig}}
\end{figure}

Figure \ref{dvi.fig} shows the intrinsic linewidth of NH\3 in the
central 10 pc of the Galaxy.  The linewidths are in color while a 6~cm
continuum emission image \citep{yus87} is overlaid in contours.  The
continuum map shows the outer edge of the expanding SNR, Sgr A East,
as well as the ionized arms of the mini-spiral (Sgr A West).  Emission
associated with the $2.6\times10^6$~M$_\odot$ supermassive black hole
\citep{eck97,ghe98}, known as Sgr A*, is seen as the point source at
the center of the map.  Previous NH\3(1,1) observations of the
Galactic Center find $\langle\dvoa\rangle\sim30$~\kms ~for the region
\citep{coi99,coi00}, similar to our original results.  However, other
molecular tracers including NH\3(3,3) show significantly smaller
linewidths \citep{gus80,arm85}.  Although the large linewidths in NH\3
(1,1) are attributed to the nearby hyperfine components, no previous
attempt was made to recover the values of the intrinsic linewidths and
subsequent calculations of the opacity did not take $\dvi$ into
account.  In Figure \ref{dvi.fig}, we have applied our new
method, and the average value of NH\3(1,1) intrinsic linewidth is
$\sim15$~\kms, which is in agreement with that of NH\3(3,3) and
results from other tracers.  {\it Therefore, we find NH\3(1,1) has
intrinsic linewidths between 10 and 20 \kms ~in the central 10 pc of
the Galaxy.}

At first glance, the NH\3(1,1) linewidth appears similar throughout
the map.  However, there are many details of interest which merit
discussion.  Readers are referred to \citet{mcg01} for a more detailed
explanation of the locations and morphologies of the features
discussed below.  Figure \ref{dvi.fig} shows that most of the gas with
$\dvi>20$~\kms ~is located along the edge of Sgr A East.  This shell
is expanding into the surrounding molecular material with an energy of
$10^{52}$ ergs, more than an order of magnitude greater than typical
supernova remnants \citep{mez89,gen90}.  Recent Chandra observations
show a central concentration in the thermal x-rays within the radio
shell suggesting that Sgr A East may be an example of a Type II metal
rich ``mixed-morphology'' supernova remnant \citep{mae01}.  Emission
from the western streamer located on the western edge of Sgr A East at
($\Delta\alpha=-60''$, $\Delta\delta=15''$) and ($-70''$,$-60$)
\citep{mcg01} has a typical intrinsic linewidth of 25 \kms,
significantly above the average for the region.  The 50 \kms ~giant
molecular cloud (GMC), also known as M --0.03,--0.07 \citep{gus81}, is
located on the northeastern edge of Sgr A East at ($100''$,$50''$) and
has $\langle\dvi\rangle\approx20$~\kms.  Finally, emission at
($110''$,$-20''$) shows $\dvi$ as large as 40 \kms.  These increased
linewidths appear to result from the impact of Sgr A East and are a
strong indication that these features are in close proximity to the
expanding shell \citep{gen90,ho91,ser92,zyl99,coi00,mcg01}.  The only
other cloud with large $\dvi$ is located at ($50''$,$0''$) and is less
than 2~pc in projected distance from the Galactic Center (assuming
R$_\odot=8.0\pm0.5$~kpc \citep{rei93}).  The large intrinsic
linewidths in this feature may be due to interactions between multiple
clouds or expansion of the front or back of Sgr A East through the
cloud.

The majority of the NH\3(1,1) emission is located to the southeast of
Sgr A* and composes the ``molecular ridge'' that connects the nearby
20~\kms ~GMC in the south (M --0.13, --0.08; \citet{gus81}) to the 50
\kms ~GMC \citep{ho91,coi99,coi00,mcg01}.  Previous observations of
the southern streamer (from ($0''$,$-150''$) to ($35''$,$-75''$))
showed an increase in the linewidth as the gas approached the nucleus
\citep{coi99,coi00}.  After disentangling opacity and linewidth we
still see the largest linewidths in the northern edge of this cloud
($30''$,$-60''$).  The large linewidth combined with high temperatures
observed in this feature \citep{coi99} may indicate that the gas is
approaching the nucleus.  Increased intrinsic linewidths in the
southern parts of these clouds may be the result of the impact of a
second supernova remnant centered at ($80''$,$-120''$) to the south
\citep{coi00}.

\begin{figure}
\plotone{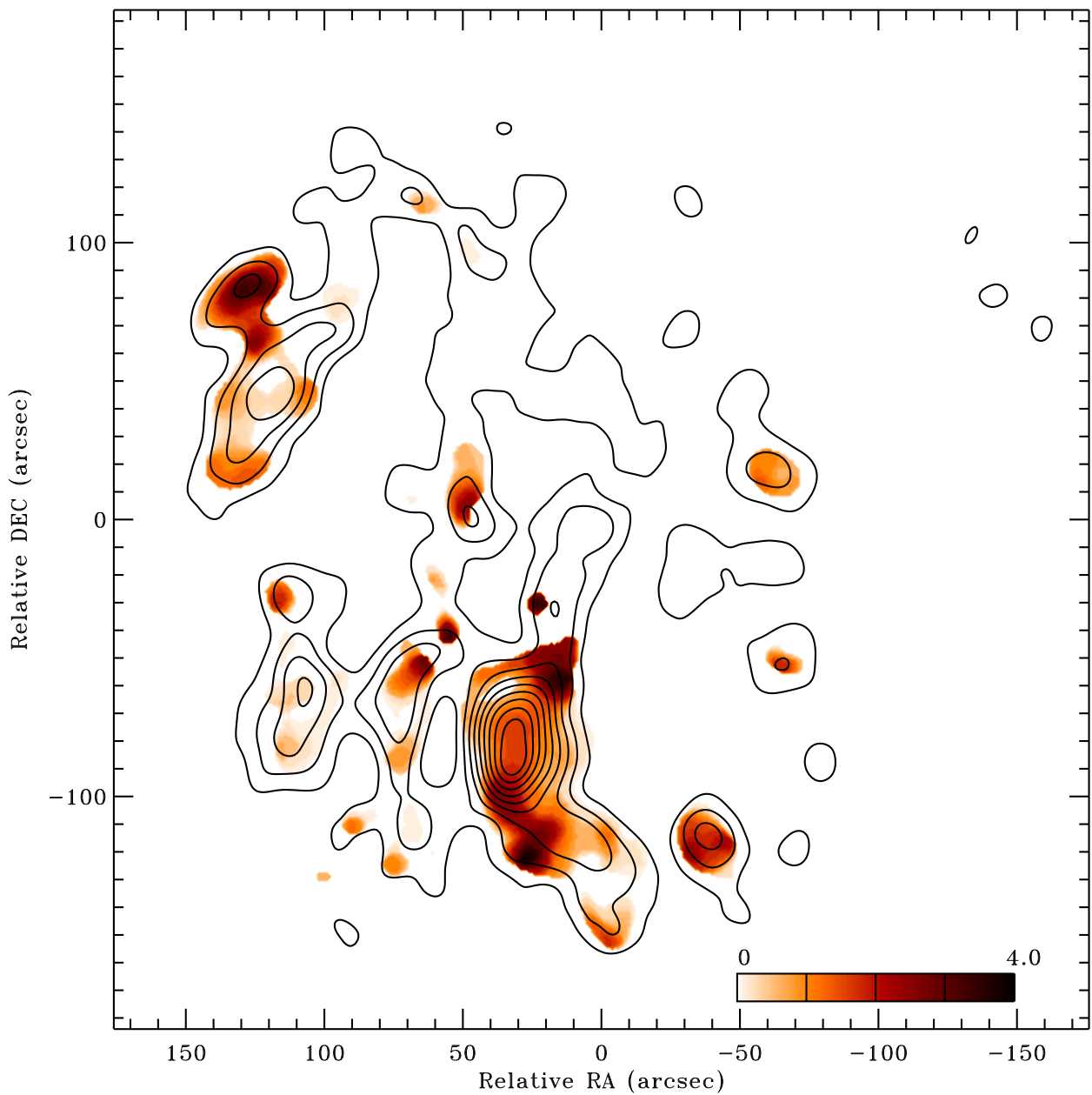}
\caption{Best estimate of $\ta$ in color overlaid on velocity
integrated NH\3(1,1) emission in 10\% contours \citep{mcg01}.  The
opacity map has been smoothed with an eight pixel median filter (see
\S \ref{gc}).  The highest opacity is associated with the southern
streamer and the 20 and 50 \kms ~GMCs while most other features have
$\ta<1.0$.  \label{11tau.fig}}
\end{figure}

When solving for the opacity in the traditional method discussed in \S
\ref{corr} we found most gas in our image appeared to have $1<\ta<2$.
Although this result is in agreement with \citet{coi99,coi00}, it is
also contrary to our assumption that all rotation inversion
transitions of NH\3 are typically optically thin at the Galactic
Center.  Figure \ref{11tau.fig} shows our result for $\ta$ in color
overlaid on 10\% contours of NH\3(1,1) velocity integrated emission
from \citet{mcg01}.  After taking the intrinsic linewidth into account
using the method described above, the opacity is no longer biased high
by the large observed linewidths and $\langle\ta\rangle=0.6$.  The
opacity is highest along the center of the southern streamer (from
($0''$,$-150''$) to ($35''$,$-75''$)).  As seen in Figure
\ref{11tau.fig}, the southern streamer is the brightest object in our
NH\3(1,1) velocity integrated image and is expected to contain
$\sim5\times10^4~\rm{M}_\odot$ of material \citep{coi99}.  This
streamer extends northwards from the 20 \kms ~GMC towards the nucleus
\citep{ho91, coi99, coi00}.  A second extension of the 20~\kms ~GMC at
($-40$,$-120$) has $\ta\approx1$.  The opacity is also elevated in the
most northeasterly clump of the 50 \kms ~GMC.  This clump is actually
associated with the center of the GMC (to see the full extent of this
cloud see \citet{den93}), but is located at the edge of our mosaic
where the gain goes to zero (see \citet{mcg01}).  Therefore, the most
massive clouds in our maps show the largest opacities as expected.

With these improved estimates of opacity and intrinsic linewidth,
the column density and mass could be estimated for each cloud.
However, calculation of $N_{(1,1)}$ requires that the peak antenna
temperature of the main line, $\Delta T_A(1,1)$, be measured for each
cloud.  In addition, to convert $N_{(1,1)}$ to $N_{\rm{NH}_3}$ the
rotational temperature of the gas must be known.  Because the
objective of this paper is to model the effect of large intrinsic
linewidths on the line profile and outline the proper method for
determining $\dvi$ and $\tm$, these quantities are not presented in
this paper.  We will present the calculated $N_{\rm{NH}_3}$,
$N_{\rm{H}_2}$, and cloud masses in a subsequent paper focused on the
overall environment in the central 10~pc of the Galaxy. For
completeness, we present a brief summary of the method to calculate
$N_{(1,1)}$ as well as the average expected value for $N_{(1,1)}$ in
this paper.

The excitation temperature, $T_{ex}$, of the gas can be calculated
using 
\bq 
T_{ex}=2.7~\rm{K} + \frac{\Delta T_A(1,1)}{1-e^{-\ta}} ~.
\eq
\noindent This equation assumes a filling factor of 1 and thus
produces a lower limit for $T_{ex}$.  $T_{ex}$ can then be used to
calculate the column density of the main transition using
\bq
\ta=\frac{c^2 h A_{1-0} f(\nu) N_1}{8 \pi \nu k T_{ex}}
\eq
\noindent where $A_{1-0}$ is the Einstein coefficient for spontaneous
emission [$1.67\times10^{-7}$ s$^{-1}$ for NH\3(1,1)] and $f(\nu)=(4
ln 2 / \pi)^\frac{1}{2}(\Delta\nu)^{-1}$ is the line profile for a
Gaussian.  $N_1$ is the total number in the upper state and is assumed
to be equal to $N_{(1,1)}/2$ because the inversion doublet is
separated by only $\sim1$~K.  Solving for $N_{(1,1)}$, we find \bq
N_{(1,1)}=1.4\times10^{14}~\rm{cm}^{-2}~ \ta T_{ex}
\left[\frac{\dvi}{10~\rm{km~s}^{-1}}\right]~.
\label{nh}
\eq 
\noindent Therefore, we expect clouds to have NH\3(1,1) column
densities on the order of $10^{14}$~cm$^{-2}$ at the Galactic Center.
In this paper, we find that $\langle\ta\rangle=0.6$ and
$\langle\dvi\rangle=15$~\kms.  Assuming a lower limit of 7~K for the
excitation temperature \citep{coi99,coi00}, we find a lower limit of
$N_{(1,1)}=8.8\times10^{14}$~cm$^{-2}$.  Our column density estimate
is approximately a factor of five smaller than the typical results of
\citet{coi99,coi00}, due mostly to the fact that $\langle\ta\rangle$
is smaller throughout the region when the effects of large intrinsic
linewidths are removed.  As long as the gas is optically thin, the
effect of the filling factor on the measured column density is
minimized.  In this case, small clumps of gas with a large column
density is essentially equivalent to gas with a larger filling factor
and a smaller column density.  

\begin{figure}
\plotone{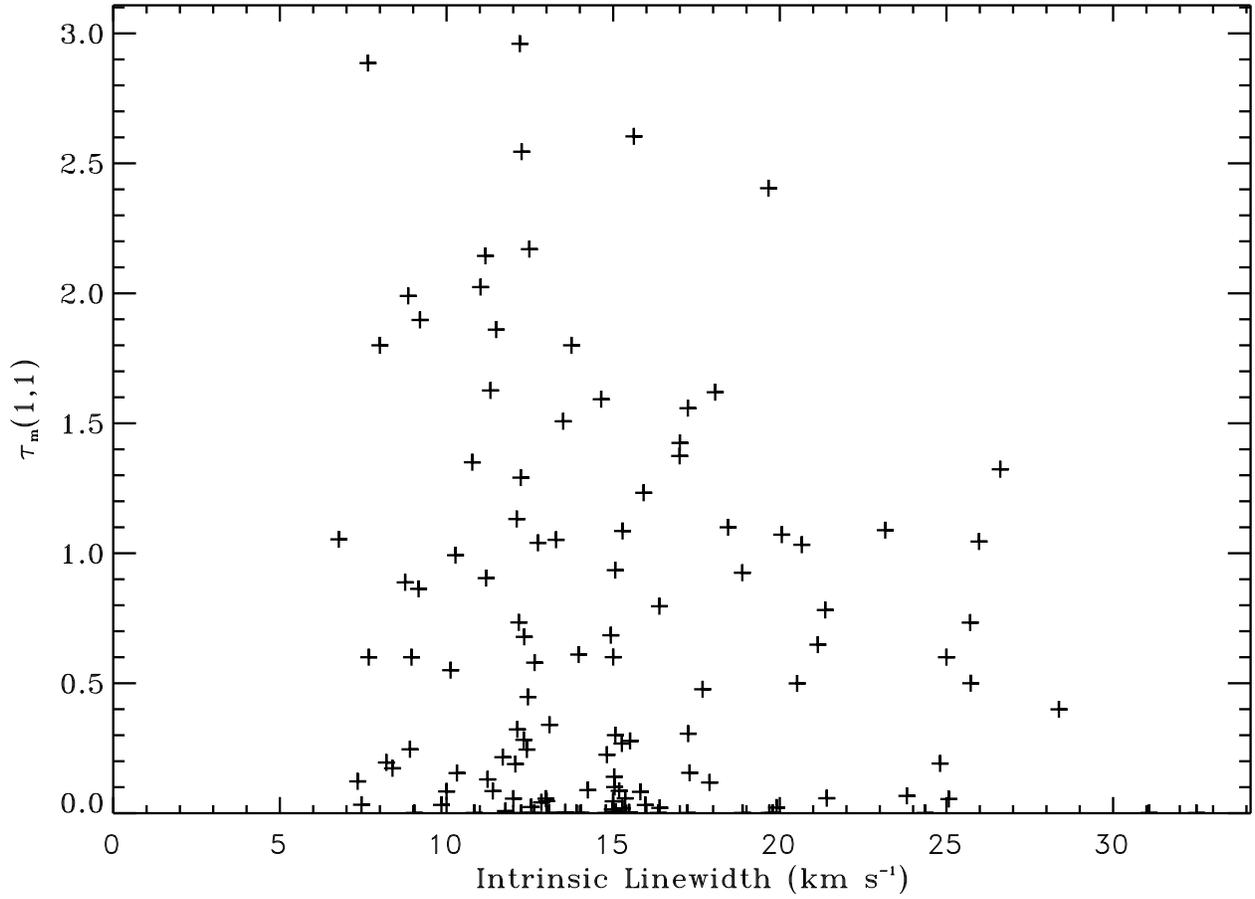}
\caption{Intrinsic linewidth, $\dvi$, as a function of
$\ta$ after application of our correction method.  Averages over the
15$''$ beam have been taken as in Figure \ref{corr.fig}$b$ and the
correlation coefficient is only -0.14. \label{fincorr.fig}}
\end{figure}

Figure \ref{fincorr.fig} shows that the correlation between $\ta$ and
$\dvi$ after the application of our method resembles a scatter plot.
The correlation coefficient is --0.14, compared to 0.69 in Figure
\ref{corr.fig}$b$.  By considering the effect of blending, we have
successfully modeled the observed linewidth and calculated accurate
intrinsic linewidths and opacities.  Finally, it is important to note
that the spread in $\dvi$ is very similar to the spread in $\dvoc$.
Therefore, one may assume that observations of NH\3(3,3) give a good
indication of the intrinsic linewidth of the gas and corrections need
not be applied in general.

\section{Conclusion \label{conc}}

In gas with large intrinsic linewidths, the five quadrupole hyperfine
components of the NH\3 rotation inversion transitions will blend into
a single-line profile.  The observed linewidth will greatly
overestimate $\dvi$ and bias the opacity to high values.  We model the
effect of $\dvi$ on observed NH\3(1,1), (2,2), and (3,3) linewidths
for a range of opacities.  A detailed method for recovery of the
intrinsic linewidth and opacity using $\dvo$ and the intensity of two
NH\3 rotation inversion transitions is then presented.  We focus on
the case where only the observed linewidth of the profile is known
which makes the method particularly useful for correcting historical
data for which the exact line profiles are not readily available.
Application of this method to our data from the central 10 pc of the
Galaxy results in the first independent measures of the opacity and
intrinsic linewidth of NH\3(1,1) near the nucleus.  The calculated
intrinsic linewidths when blending is accounted for are reduced by a
factor of $\sim2$ from the initial observed linewidths to
$\langle\dvi\rangle=15$~\kms ~and now agree with the other molecular
tracers in the region, including NH\3(3,3).  Gas along the edge of
Sgr A East or in close projected distance to the nucleus appears to
have increased intrinsic linewidths.  All except the most massive
clouds in our images show $\ta<1$.  This confirms the observation that
even the lowest energy transitions are less affected by
self-absorption than HCN(1-0) and HCO$^+$(1-0).

\acknowledgements{We would like to thank J. Herrnstein and Q. Zhang
for helpful comments.  RSM is supported in part by a Harvard
University Merit Fellowship.}

\newpage
\begin{deluxetable}{l l r r l r l } 
\tabletypesize{\small}
\tablewidth{0pt}
\tablecaption{Parameters for NH\3 Rotation Inversion Transitions \label{table}}
\tablehead{ 
\colhead{} & \colhead{} & \colhead{} & \multicolumn{2}{c}{\underline{Inner Hyperfine}} & \multicolumn{2}{c}{\underline{Outer Hyperfine}} \\
\colhead{Transition} & \colhead{Frequency}  & \colhead{Energy} & \colhead{$\Delta$v} & \colhead{Theoretical} & \colhead{$\Delta$v} & \colhead{Theoretical} \\
\colhead{}  & \colhead{(GHz)} &\colhead{K}   & \colhead{(\kms)} & \colhead{Intensity\tablenotemark{\dagger}} & \colhead{(\kms)} & \colhead{Intensity\tablenotemark{\dagger}} }
\startdata
NH\3(1,1) & 23.694495 & 23.4  & 7.7  & 0.278  & 19.3 & 0.222  \\
NH\3(2,2) & 23.722633 & 64.9  & 16.5 & 0.0628 & 25.7 & 0.0652 \\
NH\3(3,3) & 23.870129 & 124.5 & 21.4 & 0.0296 & 28.8 & 0.0300 \\
\tablenotetext{\dagger}{Intensity relative to the main line}
\enddata
\end{deluxetable}

\newpage

\newpage

\newpage

\begin{thebibliography}{dummy}
\bibitem[Armstrong \& Barrett(1985)]{arm85} Armstrong, J.T. \& Barrett, A.H. 1985, \apjs, 57, 535

\bibitem[Barranco \& Goodman(1998)]{bar98} Barranco, J.A. \& Goodman, A.A. 1998, \apj, 504, 207

\bibitem[Coil \& Ho(1999)]{coi99} Coil, A.L. \& Ho, P.T.P. 1999, \apj, 513, 752
\bibitem[Coil \& Ho(2000)]{coi00} Coil, A.L. \& Ho, P.T.P. 2000, \apj, 533, 245

\bibitem[Dent et al.(1993)]{den93} Dent, W.R.F, Matthews, H.E., Wade, R. , \& Duncan, W.D. 1993, \apj, 410, 650

\bibitem[Eckart \& Genzel(1997)]{eck97} Eckart, A. \& Genzel, R. 1997, \mnras, 284, 576

\bibitem[Genzel et al.(1990)]{gen90} Genzel, R., Stacey, G.J., Harris, A.I., Geis, N., Graf, U.U., Poglitsch, A. , \& Sutzki, J. 1990, \apj, 356, 160
\bibitem[Genzel \& Townes(1987)]{gen87} Genzel, R. \& Townes, C.H. 1987, \araa, 25, 377

\bibitem[Ghez et al.(1998)]{ghe98} Ghez, A.M., Klein, B.L., Morris, M. , \& Becklin, E.E. 1998, \apj, 509, 678

\bibitem[G\"{u}sten, Walmsley, \& Pauls(1981)]{gus81} G\"{u}sten, R., Walmsley, C.M. , \& Pauls, T. 1981, \aap, 103, 197
\bibitem[G\"{u}sten \& Downes(1980)]{gus80} G\"{u}sten, R. \& Downes, D. 1980, \aap, 87, 6

\bibitem[Ho(1977)]{ho77t} Ho, P.T.P. 1977, Ph.D. thesis, Massachusetts Institute of Technology
\bibitem[Ho et al.(1991)]{ho91} Ho, P.T.P., Ho, L.C., Szczepanski, J.C., Jackson, J.M., Armstrong, J.T. , \& Barrett, A.H. 1991, \nat, 350, 309
\bibitem[Ho et al.(1977)]{ho77} Ho, P.T.P., Martin, R.N., Myers, P.C., \& Barrett, A.H. 1977, \apjl, 215, 29
\bibitem[Ho et al.(1990)]{ho90} Ho, P.T.P., Martin, R.N., Turner, J.L., \& Jackson, J.M. 1990, \apjl, 355, 19
\bibitem[Ho \& Townes(1983)]{ho83} Ho, P.T.P. \& Townes, C.H. 1983, \araa, 21, 239

\bibitem[Maeda et al.(2001)]{mae01}Maeda, T. et al. 2001, astro-ph/0102183

\bibitem[McGary et al.(2001)]{mcg01} McGary, R.S., Ho, P.T.P., \& Coil, A.L. 2001, \apj, 559, 326

\bibitem[Mezger et al.(1989)]{mez89} Mezger, P.G., Zylka, R. Salter, C.J., Wink, J.E., Chini, R. Kreysa, E. , \& Tuffs, R. 1989, \aap, 209, 337

\bibitem[Reid(1993)]{rei93} Reid, M.J. 1993, \araa, 31, 345

\bibitem[Serabyn, Lacy, \& Achtermann(1992)]{ser92} Serabyn, E., Lacy, J.H. , \& Achermann, J.M. 1992, \apj, 395, 166

\bibitem[Townes \& Schalow(1975)]{tow75} Townes, C.H. \& Schalow, A.L. 1975, {\it Microwave Spectroscopy}, New York:Dover

\bibitem[Zylka(1999)]{zyl99} Zylka, R., G\"{u}sten, R., Philipp, S., Ungerechts, H., Mezger, P.G. , \& Duschl, W.J. 1999, in ASP Conf. Ser, 186, The Central Parsecs of the Galaxy, ed. H. Falcke et al. (San Francisco: ASP), 415

\bibitem[Yusef-Zadeh \& Morris(1987)]{yus87} Yusef-Zadeh, F. \& Morris, M. 1987, \apj, 320, 545
\end{thebibliography}
\end{document}